\documentclass[%
 reprint,
 amsmath,amssymb,
 aps,
]{revtex4-1}

\usepackage{graphicx}
\usepackage{dcolumn}
\usepackage{bm}
\usepackage{comment}
\usepackage{color}
\usepackage{hyperref}

\begin{document}

\preprint{APS/123-QED}

\title{Emergence of inertial waves from coherent vortex source in Yukawa medium}
\author{Akanksha Gupta}
\email{akgupt@iitk.ac.in}
\affiliation{%
Indian Institute of Technology Kanpur, Kanpur-208016, India
 }%
 \author{Rajaraman Ganesh}%
 \email{ganesh@ipr.res.in}

\affiliation{%
 Institute for Plasma Research, HBNI, Bhat, Gandhinagar - 382428, India
 }%

\begin{abstract}
\noindent The  evolution of isotropic, nondispersive, inertial wave, emerging from an unsteady initial coherent vortex is studied for strongly correlated Yukawa medium  using 2D molecular dynamics simulation. In this study, the effect of azimuthal speed of vortex source, strong correlation, large screening and compressibility of the medium  over the propagation of generated  inertial wave have been presented. It has been observed that these  inertial waves only exist when the  speed of vortex source ($U_{0}$) is larger or equal to the longitudinal sound speed of the system. Estimated speed of nonlinear wave $(C_{NLW})$ is found to be always larger than the  transverse sound speed ($C_{t}$)  of the system for the range of coupling and screening parameters. In this study, we find that spontaneously generated  nonlinear inertial wave speed in Yukawa medium is suppressed by compressibility and dust-neutral drag of the system and is less sensitive to coupling strength. A transition
from incompressible  to
compressible  Yukawa liquid is observed. This transition depends on the screening parameter and azimuthal speed of vortex source.  Existence of a critical  Mach number $M_{c}\approx0.35$ is found above which nolinear wave is found to exists, indicating compressible nature of the medium.
\end{abstract}

\pacs{Valid PACS appear here}
\maketitle

\section{Introduction}

\noindent  Grain medium (plasma with micron and sub-micron sized dust grains), also known as complex or dusty plasma which behaves like viscoelastic medium, facilitate linear and nonlinear waves \cite{kaw,PielPRL2002}. Such grain medium  can behave like viscous, visco-elastic and elastic medium and
can be characterized by two  dimensionless parameters, screening parameter $\kappa=a/\lambda_{D}$ (where $a$ is inter-grain-spacing and $\lambda_{D}=\lambda_{i}\lambda_{e}/\sqrt{\lambda_{i}^{2}+\lambda_{e}^{2}}$ is  Debye length of background plasma, $\lambda_{i}$, $\lambda_{e}$ are Debye length of electron and ion respectively) and coupling parameter $\Gamma = Q_{d}^{2}/(4\pi\varepsilon_{0}ak_{B}T_{d} )$ wherein $Q_{d}$  and $T_{d}$ are charge and temperature  of grain \citep{morfillRMP}. Such plasma occur in nature and in laboratory as well for example, comets, planetary rings, white dwarf, earth's atmosphere and in plasma processing reactors, plasma torch and fusion devices \citep{fusion}. \\

\noindent  Due to strong coupling nature of the medium both longitudinal and transverse  wave modes may be expect to be correlated in the grain medium. Longitudinal modes exist through all state of dusty plasma, however Transverse mode occur due to finite elasticity of the medium and hence exist in liquid and solid regime.  Transverse shear waves, are also known as  surface waves and studied theoretically \cite{kaw}, numerically \cite{vikram2014} and experimentally \cite{piel2006laser} in the strongly correlated grain medium (or complex plasma).   In the past, various wave related phenomena have been studied in strongly correlated grain medium for example, compressional and shear
modes \citep{nunomura2000transverse,nosenko2002nonlinear},  Mach cones \citep{nosenko2002observation}, transverse waves \citep{Pramanik} and driven transverse wave \cite{bandyopadhyay2008driven}. In the past,  using molecular dynamics simulation and experiments the radiation of elastic waves in a plasma crystal using small dipole source has been observed \citep{PielPRL2002}.\\

\noindent There are many bodies in our solar system which have solid rotating inner core and a fluid outer core \cite{zhang_earnshaw_liao_busse_2001}.   Inertial waves are found to  emerge from a localized rotating  inner core, for example, these  waves exist at the outer core of Earth because of the Earth's rotation\cite{ Nature_1987,book:greenspan1968}. In such case, restoring force for inertial waves is the Coriolis force. Understanding of such wave propagation has many important applications in geophysics, geodynamo, and Earth's core dynamics.  In present study, we use a Yukawa liquid (dusty plasma) as a prototype  or a visco-elastic medium to study such hydrodynamical waves, wherein restoring force is provided by the finite compressibility and elasticity of the medium. In the present study, we address several important questions,  for example, for a single value of  strong correlation ($\Gamma$) and screening ($\kappa $) is there any inertial wave generated by  the presence of coherent localized vortex?, how the  inertial wave changes its nature with an azimuthal speed of coherent localized vortex? what are the effect of variation of $\Gamma$ and $\kappa $ over the generated wave. In the present work, for the first time, using molecular dynamics simulation, we study the emergence of non-linear inertial waves due to azimuthal motion of  ideal rotational flow and  the effect of strong correlation of the medium over such waves. \\
   
\noindent Dusty or complex plasmas are often modelled by repulsive interaction potential called screened Coulomb potential or Yukawa potential $  U(\boldsymbol{r}_{i}) = Q_{d}^2/4 \pi \epsilon_0  \sum \limits_{j\neq i}^N \frac{e^{-r_{ij}/\lambda_D}}{r_{ij}}$ 
where $r_{ij}=\vert \boldsymbol{r}_{i}-\boldsymbol{r}_{j}\vert$ is the inter particle distance of $i^{th}$ and $j^{th}$ particle. We perform molecular dynamics simulation for  unbounded (infinite) system hence  periodic boundary condition and no confining external force have been used. Space is normalised to Wigner-Seitz radius and time is normalised to $\omega_{0}^{-1}$, where $\omega_{0}^{-1}=\sqrt{2}\omega_{pd}^{-1}$ ($\omega_{pd}=\sqrt{n_{0}Q_{d}^{2}/M_{d}\epsilon_{0}}$, where $n_{0}$ and $M_{d}$ are two-dimensional dust density and average mass of dust grain respectively ) \citep{thirdpaper}. We consider ambient plasma  properties to be invariant and model only grain dynamics because of slow response of grain medium as compared to electrons and ions which is due to large mass of the dust. In later part of this work, we also discuss the effect of neutral-dust collisions.  The $N$-body problem is numerically integrated using our parallelized MD code \cite{AshwinPRE}.  
\section{Flow Description}
To study the vortex flow dynamics of rotational shear flow in strongly correlated liquids, a Rankine vortex \citep{giaiotti2006rankine,loiseleux1998effect,hoff_harlander_egbers_2016} is chosen as the initial condition. Rankine vortex model is an azimuthal shear flow with two flow regions namely $(a)$ inner region of the flow $r<R$, which has a rotational profile like that of a rigid rotator  $(b)$ outer region  $r\geq R$,. The mathematical expression of Rankine velocity profile is  $\textbf{V}=v_{r}\hat{r}+v_{\theta}\hat{\theta}+v_{z}\hat{z} $, where  \[v_{r}=0, v_{\theta}(r)=
  \begin{cases}
    U_{0}r/R       & \quad \text{if } r < R\\
     U_{0}R/r  & \quad \text{if } r \geq R\\
 \end{cases},
v_{z}=0 \]  where $v_{r}$, $v_{\theta}(r)$, $v_{z}$ are radial, azimuthal and axial velocities respectively. $U_{0}$ is the strength of azimuthal velocity, $r$ and $R$ are radial coordinates and radius of Rankine vortex core respectively. In  Cartesian co-ordinate the $x$ and $y$ component of velocities are $v_{x}=v_{\theta}sin\theta$ and $v_{y}=v_{\theta}cos\theta$, where $v_{x}$ and $v_{y}$ are particle velocities in cartesian system.\\

\section{Computational Analysis and Results}

\noindent To start MD simulation  in Yukawa medium, a large number of particles  $N_{d}=62500$ were thermalized for desired value of coupling strength or inverse of temperature. Particles are thermalised using a Gaussian thermostat \cite{EVANS} for time $t=200$.  The thermostat is put off for $t=200$. To study the vortex dynamics of rotational shear flow, the Rankine vortex profile is superimposed over thermalised particles velocities. The reduced number density $n_{0}=\pi^{-1}$. We do not consider Ewald sums \cite{Salin} due to the large size of the simulation box $L_{x}=L_{y}=L = 443.12$.   To obtain macro-scale quantities for example, averaged velocity and averaged vorticity from microscopic information, we perform ``process of fluidization" \citep{thirdpaper}. \\
\begin{figure}
\includegraphics[width=3.5in,height=4.0in]{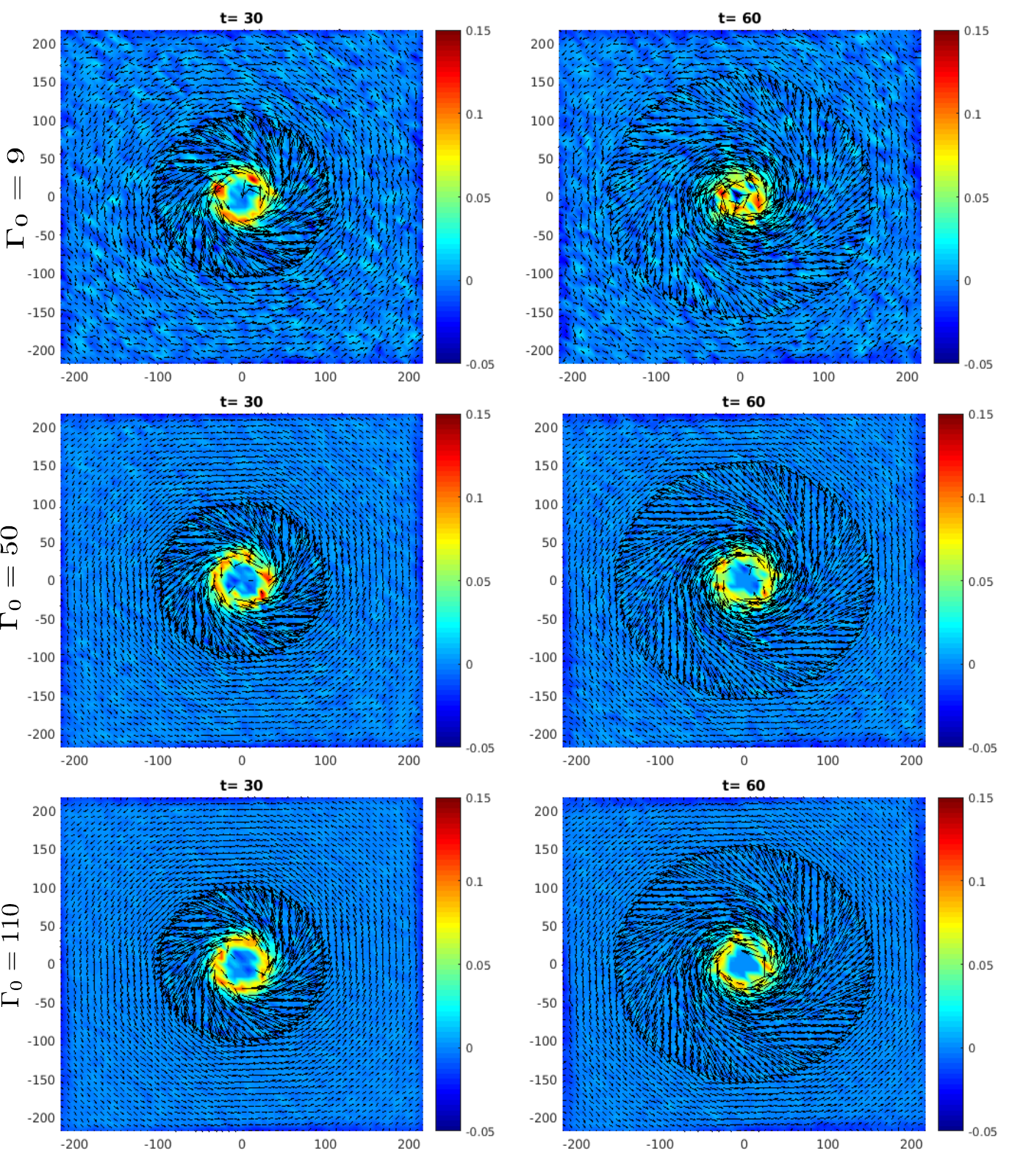}
\caption{color online: Contour plot of fluid vorticity $(\omega= \nabla \times \textbf{V})$ obtained from molecular data for initial velocity $U_{0}=5.0$, screening parameter $\kappa=1.0$, wherein black coloured arrows shows the velocity field. The radius of Rankine vortex is considered to be R=10.  The grain velocity in the bins are fluidized through a $55 \times 55$ grid to construct vorticity. }
\label{fig-vorticityevolve}
\end{figure}

\begin{figure}
\includegraphics[width=3.5in,height=2.5in]{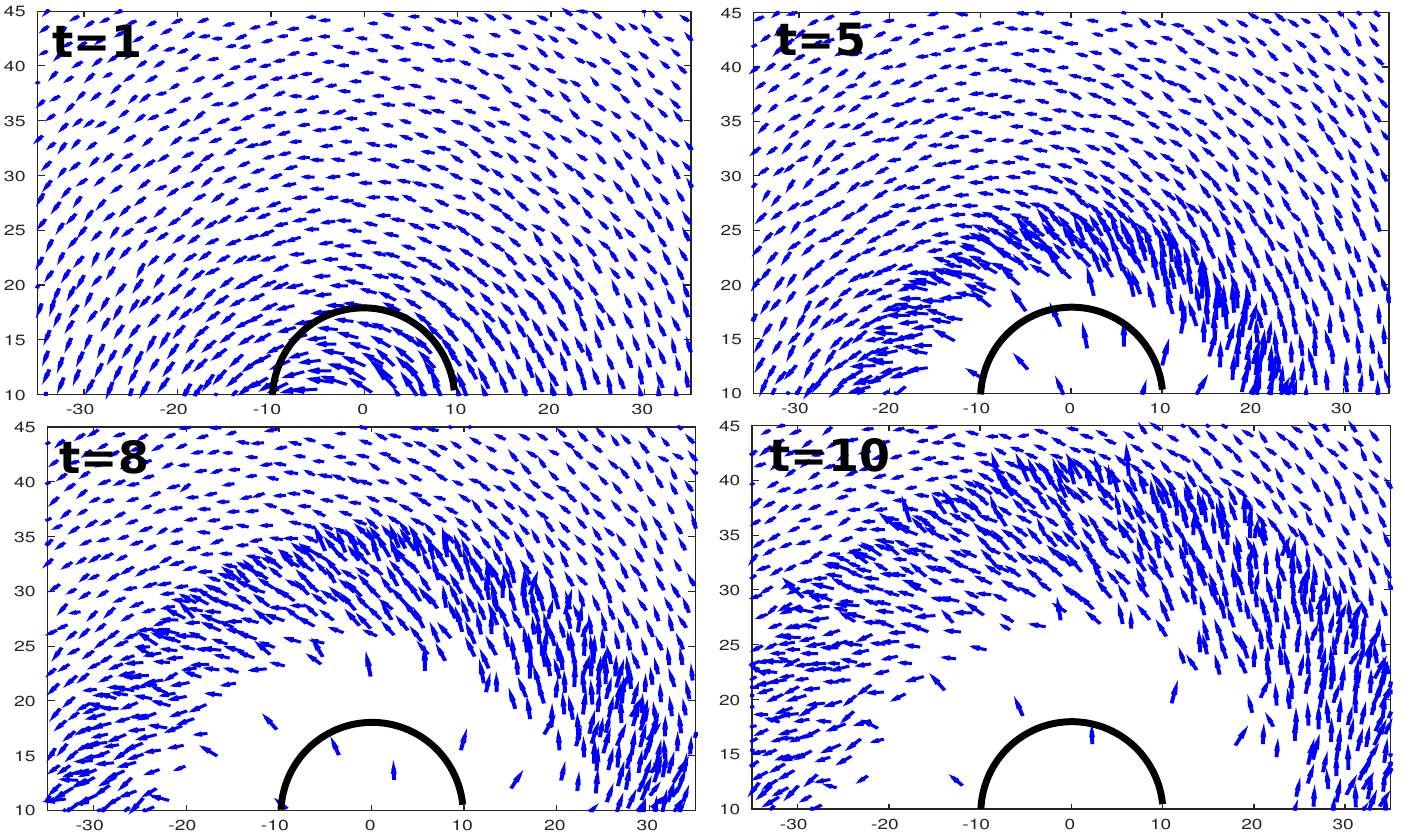}
\caption{color online: Vector plot of microscale particle velocity  for $U_{0}=5.0$, $\Gamma_{0}=50$ and $\kappa=1.0$, where velocity vectors are represented as arrows   in the range of $x$=[-35, +35] and $y$=[+10, +45]. Figure shows the transverse variation of number of particles (radially outward from the source) from the azimuthal vortex source. Inner semi-circle of radius $R=10.0$ shows the Rankine vortex region. }
\label{fig-quiver}

\end{figure}

\begin{figure*}[h!]
\includegraphics[width=6.0in,height=3.5in]{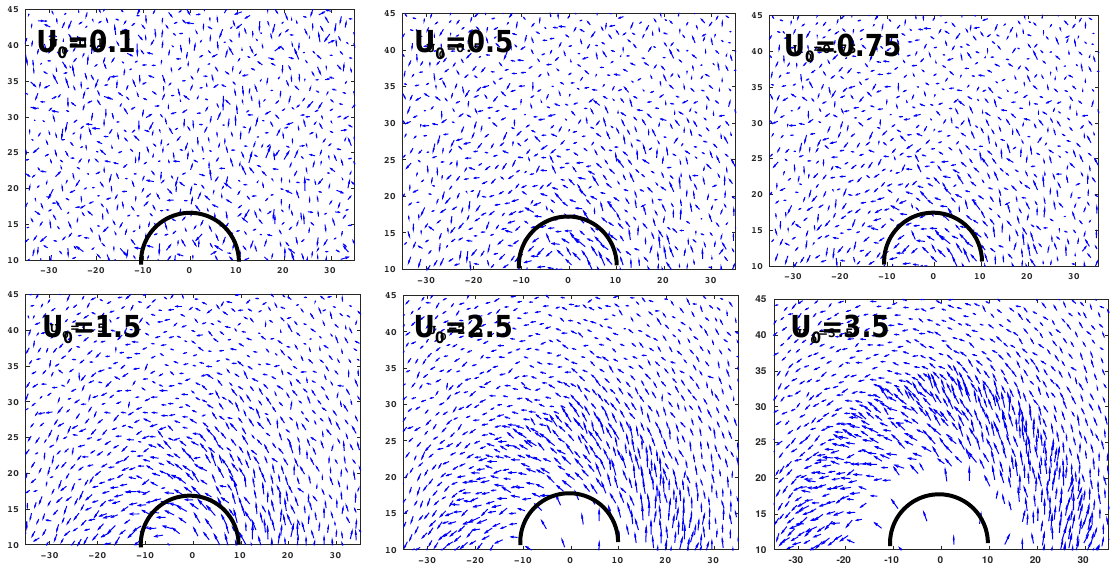}
\caption{color online:  Vector plot of microscale particle velocity  at time $t=10.0$  for various value of $U_{0}$ (= 0.1, 0.5, 0.75, 1.5, 2.5, 3.5) with $\Gamma_{0}=50$ and $\kappa=1.0$, where velocity vectors are represented as arrows  in the range of $x$=[-35, +35] and $y$=[+10, +45].}
\label{fig-particleU}
\end{figure*}

\begin{figure*}[h!]
\includegraphics[width=6.0in,height=3.5in]{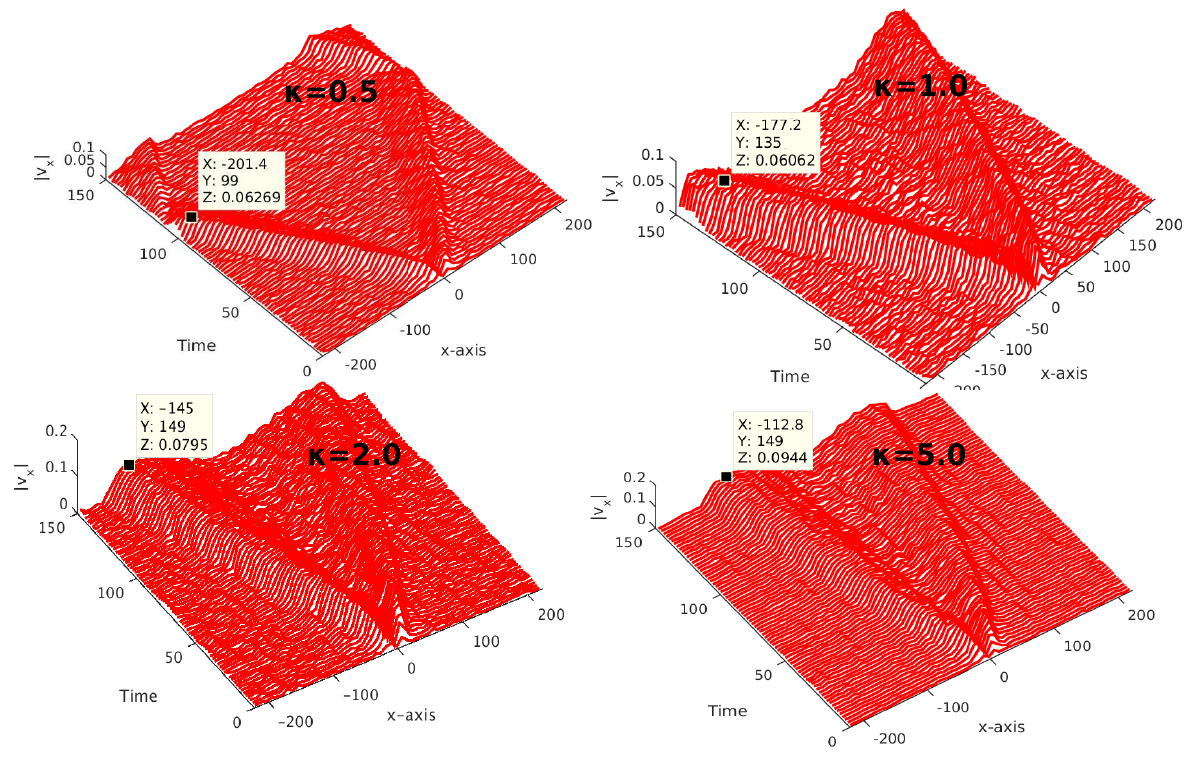}
\caption{color online: 3D plot of space, time and absolute value of fluidized $y$ averaged velocity along $x$ direction $|v_{x}(x,t)|$. Propagation of  inertial wave with various value of screening parameter $\kappa$ for $U_{0} = 2.5$ and $\Gamma_{0}= 50$. The slope
 $\Delta x/\Delta t$ gives the propagation speed of wave along $x$ direction ($C_{x}$).  $C_{y}=\Delta y/\Delta t$ is found to be same as $C_{x}$. }
\label{fig-NLW}
\end{figure*}
\noindent  Time evolution of vortex rotation and  wave propagation have been shown in Fig.\ref{fig-vorticityevolve}.  Increasing strong correlation by increasing coupling strength $\Gamma$ at constant screening parameter $\kappa=1$  does not show any significant qualitative differences in the structure and time evolution of wave as time passes. To measure the quantitative changes, we estimate the speed of nonlinear inertial wave $C_{NLW}$. Speed of emerging non-linear wave has been calculated by  $C_{NLW}=\sqrt{C_{x}^{2}+C_{y}^{2}}$, where   $C_{x}=\Delta x/\Delta t$, $C_{y}=\Delta y/\Delta t$ ($\Delta x$ and $\Delta y $ are distance traveled by wave along $x$ and $y$ directions in time $\Delta t$). It is important to note that the propagated wave is isotropic in space and hence the speed along $x$ and $y$ directions are found to be close to each other i.e $C_{x} \approx C_{y}$. For each value of initial coupling parameter $\Gamma_{0} = 9, 50, 110$, with various initial azimuthal speeds $U_{0}$ = 0.75, 2.5, 3.5, 5.0, speed of  inertial wave $C_{NLW}$  are $\approx$ 1.51, 1.81, 2.26 and 3.11 respectively. It has been observed that the increasing strength of rotational vortex ($U_{0}\uparrow$) enhances the speed of propagation of  wave ($C_{NLW}\uparrow$). For example, waves generated from vortex source of higher azimuthal velocity  touch the boundary first rather than lower one and re-appear on the other side of the boundary because of periodic boundary condition. Fig.\ref{fig-quiver} shows the particles orientation  radially outward from the centre of vortex for $U_{0}=5.0$, $\Gamma_{0}=50$ and $\kappa=1.0$. 
 Due to strong correlation between the particles of the medium, the bunch of particles near the edge of vortex source resume its natural shape after vortex rotation and therefore, nearest particles undergo shear, by this way the wave propagate in the medium. The wave propagation from the vortex source is crucially depends upon the azimuthal speed of vortex source. In Fig.\ref{fig-particleU} [details are present in the caption], it is shown that the wave propagation starts when $U_{0}\geqslant 0.75$, which is much grater than transverse sound speed $C_{t}$ and thermal speed $C_{th}=\sqrt{2/\Gamma_{0}}=0.2$. We shall come to this point later. Two sound speeds in the system exist one for compressional ($C_{l}$ towards $\hat{\theta}$ direction) and other for shear ($C_{t}$ towards $\hat{r}$ direction) waves. In the presence of macroscale vortex flows speed larger and finite compressiblity these modes get coupled with each other.   We have repeated our numerical experiments for various values of screening parameter [$\kappa=0.5-5.0$] with constant value of azimuthal speed and coupling parameter.  Fig.\ref{fig-NLW} shows that the screening parameter suppresses the speed of linear wave. It is found that in Yukawa medium for given value of $\Gamma$ and $\kappa$, sound speeds ($C_{l}$ and $C_{t}$) are mainly dependent on $\kappa$ and insensitive to $\Gamma$  \citep{khrapak2016relations}. We have calculated the $C_{l}$ and $C_{t}$ using equilibrium MD simulation for our system for $\Gamma=50,$ and  $\kappa=1$ using pair correlation related formula  \citep{khrapak2016relations}.  In present study, there are three main speed exist in Yukawa  medium i.e $C_{l}$, $C_{t}$  and $U_{0}$. In Fig.\ref{fig-NLW}, 3D plot of space, time and absolute value of fluidized $y$ averaged velocity along $x$ direction $|v_{x}(x,t)|$ has been plotted for various values of $\kappa$ for $\Gamma=50$, $U_{0}=2.5$. \\
\begin{figure}[h!]
\includegraphics[width=3.5in,height=3.0in]{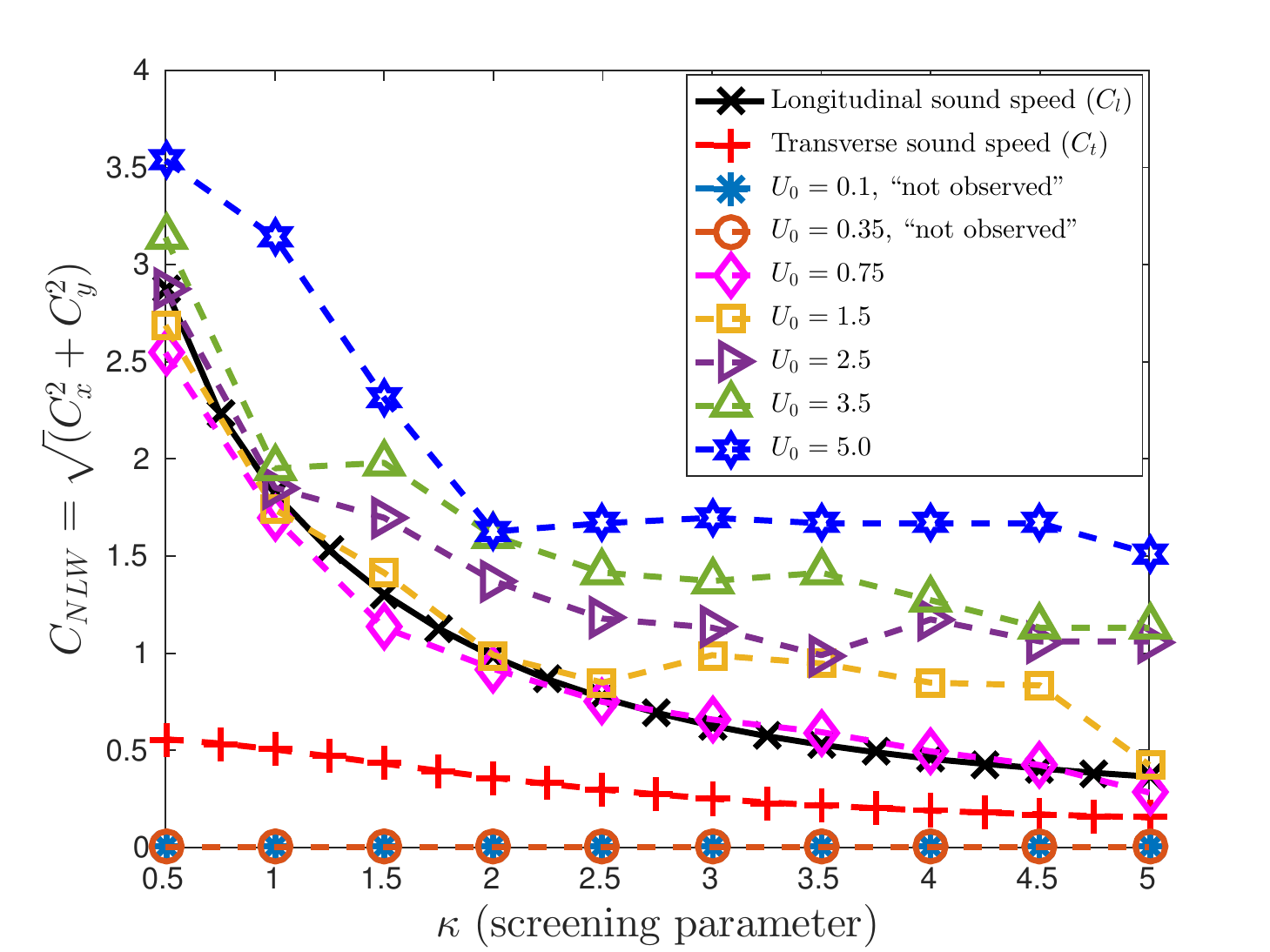}
\caption{color online:  Speed of  inertial wave $C_{NLW}$ with screening parameter for coupling parameter $\Gamma_{0}=50$ and various values of equilibrium velocity $U_{0}$. }
\label{fig_NLWspeed}
\end{figure}
\begin{figure}[h!]
\includegraphics[width=3.5in,height=3.0in]{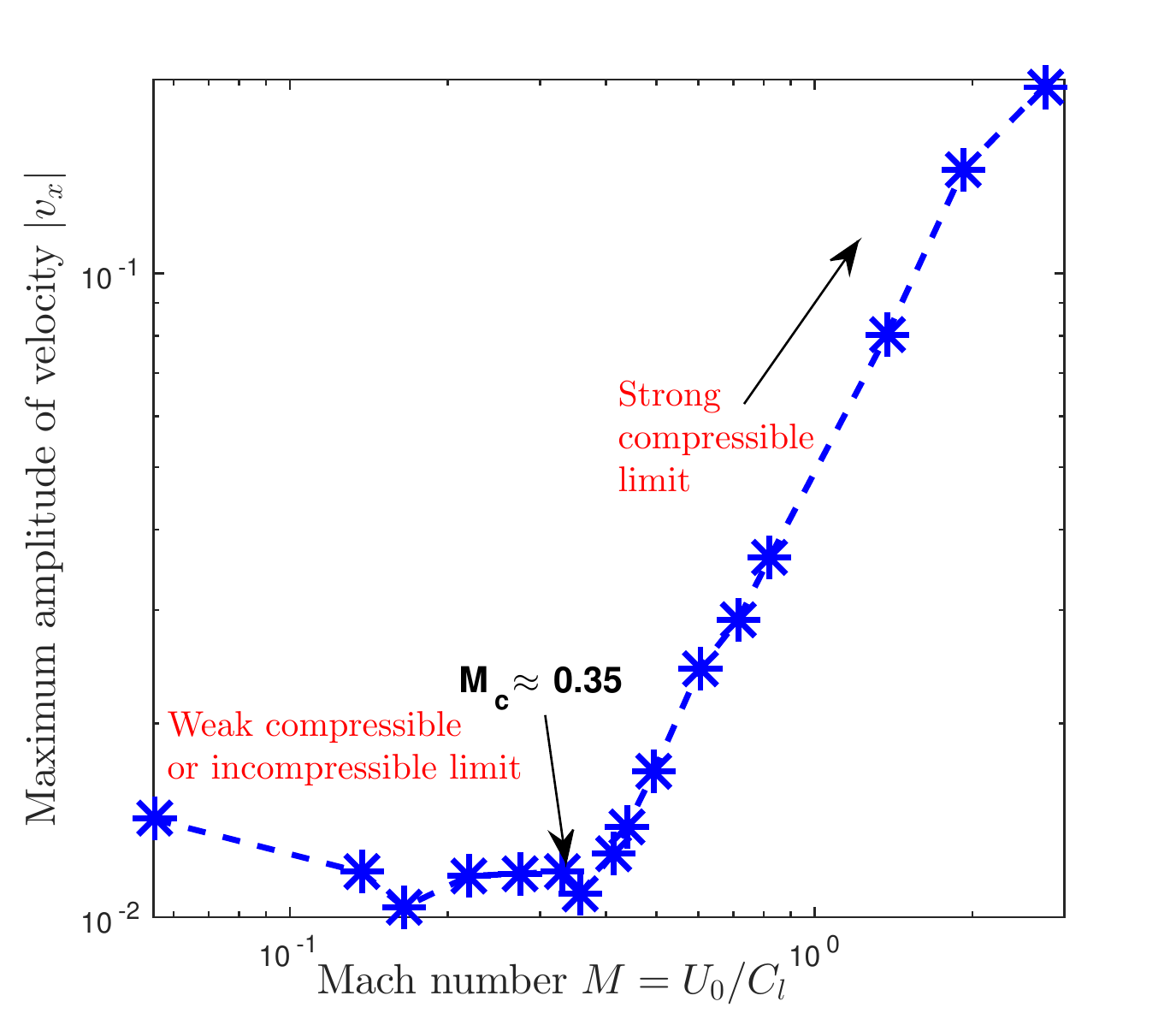}
\caption{color online: Maximum amplitude of absolute value of nonlinear wave velocity along $x$ direction ($v_{x}$) with Mach number $M=U_{0}/C_{l}$ for $\Gamma_{0}=50$ and $\kappa=1.0$. For our normalisation $C_{l}=1.82$. Figure shows the existence of a critical value of Mach number $M_{c}$ above which nolinear wave has been observed. 
}
\label{fig_machvary}
\end{figure}

\noindent The slope $\Delta x/\Delta t$ and $\Delta y/\Delta t$ give the propagation speed of wave along $x$  ($C_{x}$) and $y$ $(C_{y})$ directions respectively. Fig.\ref{fig-NLW} clearly show that screening parameter ($\kappa \uparrow$) decreases the speed of wave and increases the amplitude of velocity [see $z$ axis of Fig.\ref{fig-NLW}]. It is observed that the  inertial wave propagate when $C_{t}<U_{0}\leq C_{l}$ and $C_{t}< C_{l}<<U_{0}$ and no wave has been observed when $U_{0}\leq C_{t}<C_{l}$. Fig.\ref{fig_NLWspeed} shows the speed of linear wave with screening parameter as $\kappa$ decreases the sound speed of the system and makes the system (or medium) to be more compressible. In this study, we find that spontaneously generated rotational wave speed in Yukawa medium is suppressed by compressibility and independent of coupling strength. Emergent  wave is  isotropic and non-dispersive.  In fluid dynamics, it is well known and experimentally observed that the fluid medium below Mach number $M<0.3$ is incompressible or weakly compressible \cite{Kundu:book}. In our work, Mach number variation (via increasing initial velocity magnitude $U_{0} \uparrow$) study with maximum amplitude of absolute value of  inertial wave velocity along $x$ direction ($v_{x}$)  for $\Gamma=50$ and $\kappa=1.0$, a critical value of Mach number $M_{c}\approx 0.35$ is observed above which the medium gets significant compressibility to sustain the  wave [see Fig.\ref{fig_machvary} ].    We also studied the generation of wave in presence of neutral-dust collision by incorporating the neutral drag force in the equation of motion. It is found that the neutral drag decreases $C_{NLW}$ values. From our simulations, for $\Gamma_{0}=50$, $\kappa=1$, $U_{0}=5.0$  parameters with realistic  neutral-drag coefficients $\nu_{d}= 1.0\times 10^{-3}$,  $ 4.0\times 10^{-3}$,  $8.0\times 10^{-3}$ the wave speed $C_{NLW} \approx 2.76$, 2.47, 2.26.\\

In present work, we observe isotropic and non-dispersive wave  emerged from
a localized source in strongly correlated dusty plamsa which also behave as a viscoelastic medium. We  studied for the first time, the effect of azimuthal speed of vortex source, strong correlation, large screening and compressibility of the medium  over the propagation of generated  wave using non-equilibrium molecular dynamics simulation. We have also observed the incompressible (water-like) to compressible flow transition via increasing value of initial velocity magnitude $U_{0}$. We find that spontaneously generated  wave speed in Yukawa medium is suppressed by compressibility and dust-neutral drag of the system and is less sensitive to coupling strength. 

\section{Acknowledgement}
All simulations have been performed at  Uday and Udbhav clusters at Institute for Plasma Research and HPC2013-IITK cluster of IIT Kanpur.  


\nocite{}


\begin{thebibliography}{23}%
\makeatletter
\providecommand \@ifxundefined [1]{%
 \@ifx{#1\undefined}
}%
\providecommand \@ifnum [1]{%
 \ifnum #1\expandafter \@firstoftwo
 \else \expandafter \@secondoftwo
 \fi
}%
\providecommand \@ifx [1]{%
 \ifx #1\expandafter \@firstoftwo
 \else \expandafter \@secondoftwo
 \fi
}%
\providecommand \natexlab [1]{#1}%
\providecommand \enquote  [1]{``#1''}%
\providecommand \bibnamefont  [1]{#1}%
\providecommand \bibfnamefont [1]{#1}%
\providecommand \citenamefont [1]{#1}%
\providecommand \href@noop [0]{\@secondoftwo}%
\providecommand \href [0]{\begingroup \@sanitize@url \@href}%
\providecommand \@href[1]{\@@startlink{#1}\@@href}%
\providecommand \@@href[1]{\endgroup#1\@@endlink}%
\providecommand \@sanitize@url [0]{\catcode `\\12\catcode `\$12\catcode
  `\&12\catcode `\#12\catcode `\^12\catcode `\_12\catcode `\%12\relax}%
\providecommand \@@startlink[1]{}%
\providecommand \@@endlink[0]{}%
\providecommand \url  [0]{\begingroup\@sanitize@url \@url }%
\providecommand \@url [1]{\endgroup\@href {#1}{\urlprefix }}%
\providecommand \urlprefix  [0]{URL }%
\providecommand \Eprint [0]{\href }%
\providecommand \doibase [0]{http://dx.doi.org/}%
\providecommand \selectlanguage [0]{\@gobble}%
\providecommand \bibinfo  [0]{\@secondoftwo}%
\providecommand \bibfield  [0]{\@secondoftwo}%
\providecommand \translation [1]{[#1]}%
\providecommand \BibitemOpen [0]{}%
\providecommand \bibitemStop [0]{}%
\providecommand \bibitemNoStop [0]{.\EOS\space}%
\providecommand \EOS [0]{\spacefactor3000\relax}%
\providecommand \BibitemShut  [1]{\csname bibitem#1\endcsname}%
\let\auto@bib@innerbib\@empty
\bibitem [{\citenamefont {Kaw}\ and\ \citenamefont {Sen}(1998)}]{kaw}%
  \BibitemOpen
  \bibfield  {author} {\bibinfo {author} {\bibfnamefont {P.~K.}\ \bibnamefont
  {Kaw}}\ and\ \bibinfo {author} {\bibfnamefont {A.}~\bibnamefont {Sen}},\
  }\href@noop {} {\bibfield  {journal} {\bibinfo  {journal} {Physics of
  Plasmas}\ }\textbf {\bibinfo {volume} {5}},\ \bibinfo {pages} {(10), 3552}
  (\bibinfo {year} {1998})}\BibitemShut {NoStop}%
\bibitem [{\citenamefont {Piel}\ \emph {et~al.}(2002)\citenamefont {Piel},
  \citenamefont {Nosenko},\ and\ \citenamefont {Goree}}]{PielPRL2002}%
  \BibitemOpen
  \bibfield  {author} {\bibinfo {author} {\bibfnamefont {A.}~\bibnamefont
  {Piel}}, \bibinfo {author} {\bibfnamefont {V.}~\bibnamefont {Nosenko}}, \
  and\ \bibinfo {author} {\bibfnamefont {J.}~\bibnamefont {Goree}},\
  }\href@noop {} {\bibfield  {journal} {\bibinfo  {journal} {Phys. Rev. Lett.}\
  }\textbf {\bibinfo {volume} {89}},\ \bibinfo {pages} {085004} (\bibinfo
  {year} {2002})}\BibitemShut {NoStop}%
\bibitem [{\citenamefont {Morfill}\ and\ \citenamefont
  {Ivlev}(2009)}]{morfillRMP}%
  \BibitemOpen
  \bibfield  {author} {\bibinfo {author} {\bibfnamefont {G.~E.}\ \bibnamefont
  {Morfill}}\ and\ \bibinfo {author} {\bibfnamefont {A.~V.}\ \bibnamefont
  {Ivlev}},\ }\href@noop {} {\bibfield  {journal} {\bibinfo  {journal} {Rev.
  Mod. Phys.}\ }\textbf {\bibinfo {volume} {81}},\ \bibinfo {pages} {1353}
  (\bibinfo {year} {2009})}\BibitemShut {NoStop}%
\bibitem [{\citenamefont {de~Angelis}(2006)}]{fusion}%
  \BibitemOpen
  \bibfield  {author} {\bibinfo {author} {\bibfnamefont {U.}~\bibnamefont
  {de~Angelis}},\ }\href@noop {} {\bibfield  {journal} {\bibinfo  {journal}
  {Physics of Plasmas}\ }\textbf {\bibinfo {volume} {13}},\ \bibinfo {eid}
  {012514} (\bibinfo {year} {2006})}\BibitemShut {NoStop}%
\bibitem [{\citenamefont {Singh~Dharodi}\ \emph {et~al.}(2014)\citenamefont
  {Singh~Dharodi}, \citenamefont {Kumar~Tiwari},\ and\ \citenamefont
  {Das}}]{vikram2014}%
  \BibitemOpen
  \bibfield  {author} {\bibinfo {author} {\bibfnamefont {V.}~\bibnamefont
  {Singh~Dharodi}}, \bibinfo {author} {\bibfnamefont {S.}~\bibnamefont
  {Kumar~Tiwari}}, \ and\ \bibinfo {author} {\bibfnamefont {A.}~\bibnamefont
  {Das}},\ }\href@noop {} {\bibfield  {journal} {\bibinfo  {journal} {Physics
  of Plasmas}\ }\textbf {\bibinfo {volume} {21}},\ \bibinfo {eid} {073705}
  (\bibinfo {year} {2014})}\BibitemShut {NoStop}%
\bibitem [{\citenamefont {Piel}\ \emph {et~al.}(2006)\citenamefont {Piel},
  \citenamefont {Nosenko},\ and\ \citenamefont {Goree}}]{piel2006laser}%
  \BibitemOpen
  \bibfield  {author} {\bibinfo {author} {\bibfnamefont {A.}~\bibnamefont
  {Piel}}, \bibinfo {author} {\bibfnamefont {V.}~\bibnamefont {Nosenko}}, \
  and\ \bibinfo {author} {\bibfnamefont {J.}~\bibnamefont {Goree}},\
  }\href@noop {} {\bibfield  {journal} {\bibinfo  {journal} {Physics of
  plasmas}\ }\textbf {\bibinfo {volume} {13}},\ \bibinfo {pages} {042104}
  (\bibinfo {year} {2006})}\BibitemShut {NoStop}%
\bibitem [{\citenamefont {Nunomura}\ \emph {et~al.}(2000)\citenamefont
  {Nunomura}, \citenamefont {Samsonov},\ and\ \citenamefont
  {Goree}}]{nunomura2000transverse}%
  \BibitemOpen
  \bibfield  {author} {\bibinfo {author} {\bibfnamefont {S.}~\bibnamefont
  {Nunomura}}, \bibinfo {author} {\bibfnamefont {D.}~\bibnamefont {Samsonov}},
  \ and\ \bibinfo {author} {\bibfnamefont {J.}~\bibnamefont {Goree}},\
  }\href@noop {} {\bibfield  {journal} {\bibinfo  {journal} {Physical review
  letters}\ }\textbf {\bibinfo {volume} {84}},\ \bibinfo {pages} {5141}
  (\bibinfo {year} {2000})}\BibitemShut {NoStop}%
\bibitem [{\citenamefont {Nosenko}\ \emph
  {et~al.}(2002{\natexlab{a}})\citenamefont {Nosenko}, \citenamefont
  {Nunomura},\ and\ \citenamefont {Goree}}]{nosenko2002nonlinear}%
  \BibitemOpen
  \bibfield  {author} {\bibinfo {author} {\bibfnamefont {V.}~\bibnamefont
  {Nosenko}}, \bibinfo {author} {\bibfnamefont {S.}~\bibnamefont {Nunomura}}, \
  and\ \bibinfo {author} {\bibfnamefont {J.}~\bibnamefont {Goree}},\
  }\href@noop {} {\bibfield  {journal} {\bibinfo  {journal} {Physical review
  letters}\ }\textbf {\bibinfo {volume} {88}},\ \bibinfo {pages} {215002}
  (\bibinfo {year} {2002}{\natexlab{a}})}\BibitemShut {NoStop}%
\bibitem [{\citenamefont {Nosenko}\ \emph
  {et~al.}(2002{\natexlab{b}})\citenamefont {Nosenko}, \citenamefont {Goree},
  \citenamefont {Ma},\ and\ \citenamefont {Piel}}]{nosenko2002observation}%
  \BibitemOpen
  \bibfield  {author} {\bibinfo {author} {\bibfnamefont {V.}~\bibnamefont
  {Nosenko}}, \bibinfo {author} {\bibfnamefont {J.}~\bibnamefont {Goree}},
  \bibinfo {author} {\bibfnamefont {Z.}~\bibnamefont {Ma}}, \ and\ \bibinfo
  {author} {\bibfnamefont {A.}~\bibnamefont {Piel}},\ }\href@noop {} {\bibfield
   {journal} {\bibinfo  {journal} {Physical review letters}\ }\textbf {\bibinfo
  {volume} {88}},\ \bibinfo {pages} {135001} (\bibinfo {year}
  {2002}{\natexlab{b}})}\BibitemShut {NoStop}%
\bibitem [{\citenamefont {Pramanik}\ \emph {et~al.}(2002)\citenamefont
  {Pramanik}, \citenamefont {Prasad}, \citenamefont {Sen},\ and\ \citenamefont
  {Kaw}}]{Pramanik}%
  \BibitemOpen
  \bibfield  {author} {\bibinfo {author} {\bibfnamefont {J.}~\bibnamefont
  {Pramanik}}, \bibinfo {author} {\bibfnamefont {G.}~\bibnamefont {Prasad}},
  \bibinfo {author} {\bibfnamefont {A.}~\bibnamefont {Sen}}, \ and\ \bibinfo
  {author} {\bibfnamefont {P.~K.}\ \bibnamefont {Kaw}},\ }\href@noop {}
  {\bibfield  {journal} {\bibinfo  {journal} {Phys. Rev. Lett.}\ }\textbf
  {\bibinfo {volume} {88}},\ \bibinfo {pages} {175001} (\bibinfo {year}
  {2002})}\BibitemShut {NoStop}%
\bibitem [{\citenamefont {Bandyopadhyay}\ \emph {et~al.}(2008)\citenamefont
  {Bandyopadhyay}, \citenamefont {Prasad}, \citenamefont {Sen},\ and\
  \citenamefont {Kaw}}]{bandyopadhyay2008driven}%
  \BibitemOpen
  \bibfield  {author} {\bibinfo {author} {\bibfnamefont {P.}~\bibnamefont
  {Bandyopadhyay}}, \bibinfo {author} {\bibfnamefont {G.}~\bibnamefont
  {Prasad}}, \bibinfo {author} {\bibfnamefont {A.}~\bibnamefont {Sen}}, \ and\
  \bibinfo {author} {\bibfnamefont {P.}~\bibnamefont {Kaw}},\ }\href@noop {}
  {\bibfield  {journal} {\bibinfo  {journal} {Physics Letters A}\ }\textbf
  {\bibinfo {volume} {372}},\ \bibinfo {pages} {5467} (\bibinfo {year}
  {2008})}\BibitemShut {NoStop}%
\bibitem [{\citenamefont {Zhang}\ \emph {et~al.}(2001)\citenamefont {Zhang},
  \citenamefont {Earnshaw}, \citenamefont {Liao},\ and\ \citenamefont
  {Busse}}]{zhang_earnshaw_liao_busse_2001}%
  \BibitemOpen
  \bibfield  {author} {\bibinfo {author} {\bibfnamefont {K.}~\bibnamefont
  {Zhang}}, \bibinfo {author} {\bibfnamefont {P.}~\bibnamefont {Earnshaw}},
  \bibinfo {author} {\bibfnamefont {X.}~\bibnamefont {Liao}}, \ and\ \bibinfo
  {author} {\bibfnamefont {F.~H.}\ \bibnamefont {Busse}},\ }\href {\doibase
  10.1017/S0022112001004049} {\bibfield  {journal} {\bibinfo  {journal}
  {Journal of Fluid Mechanics}\ }\textbf {\bibinfo {volume} {437}},\ \bibinfo
  {pages} {103–119} (\bibinfo {year} {2001})}\BibitemShut {NoStop}%
\bibitem [{\citenamefont {Aldridge}(1987)}]{Nature_1987}%
  \BibitemOpen
  \bibfield  {author} {\bibinfo {author} {\bibfnamefont {L.~L.~I.}\
  \bibnamefont {Aldridge}, \bibfnamefont {K.~D.}},\ }\href@noop {} {\bibfield
  {journal} {\bibinfo  {journal} {Nature}\ }\textbf {\bibinfo {volume} {325}}
  (\bibinfo {year} {1987})}\BibitemShut {NoStop}%
\bibitem [{\citenamefont {Greenspan}(1968)}]{book:greenspan1968}%
  \BibitemOpen
  \bibfield  {author} {\bibinfo {author} {\bibfnamefont {H.~P.}\ \bibnamefont
  {Greenspan}},\ }\href@noop {} {\emph {\bibinfo {title} {The Theory of
  Rotating Fluids}}}\ (\bibinfo  {publisher} {Cambridge University Press},\
  \bibinfo {year} {1968})\ p.\ \bibinfo {pages} {327}\BibitemShut {NoStop}%
\bibitem [{\citenamefont {Gupta}\ \emph {et~al.}(2016)\citenamefont {Gupta},
  \citenamefont {Ganesh},\ and\ \citenamefont {Joy}}]{thirdpaper}%
  \BibitemOpen
  \bibfield  {author} {\bibinfo {author} {\bibfnamefont {A.}~\bibnamefont
  {Gupta}}, \bibinfo {author} {\bibfnamefont {R.}~\bibnamefont {Ganesh}}, \
  and\ \bibinfo {author} {\bibfnamefont {A.}~\bibnamefont {Joy}},\ }\href@noop
  {} {\bibfield  {journal} {\bibinfo  {journal} {Physics of Plasmas}\ }\textbf
  {\bibinfo {volume} {23}},\ \bibinfo {eid} {073706} (\bibinfo {year}
  {2016})}\BibitemShut {NoStop}%
\bibitem [{\citenamefont {Joy}\ and\ \citenamefont {Ganesh}(2009)}]{AshwinPRE}%
  \BibitemOpen
  \bibfield  {author} {\bibinfo {author} {\bibfnamefont {A.}~\bibnamefont
  {Joy}}\ and\ \bibinfo {author} {\bibfnamefont {R.}~\bibnamefont {Ganesh}},\
  }\href@noop {} {\bibfield  {journal} {\bibinfo  {journal} {Phys. Rev. E}\
  }\textbf {\bibinfo {volume} {80}},\ \bibinfo {pages} {056408} (\bibinfo
  {year} {2009})}\BibitemShut {NoStop}%
\bibitem [{\citenamefont {Giaiotti}\ and\ \citenamefont
  {Stel}(2006)}]{giaiotti2006rankine}%
  \BibitemOpen
  \bibfield  {author} {\bibinfo {author} {\bibfnamefont {D.}~\bibnamefont
  {Giaiotti}}\ and\ \bibinfo {author} {\bibfnamefont {F.}~\bibnamefont
  {Stel}},\ }\href@noop {} {\bibfield  {journal} {\bibinfo  {journal} {October.
  University of Trieste}\ } (\bibinfo {year} {2006})}\BibitemShut {NoStop}%
\bibitem [{\citenamefont {Loiseleux}\ \emph {et~al.}(1998)\citenamefont
  {Loiseleux}, \citenamefont {Chomaz},\ and\ \citenamefont
  {Huerre}}]{loiseleux1998effect}%
  \BibitemOpen
  \bibfield  {author} {\bibinfo {author} {\bibfnamefont {T.}~\bibnamefont
  {Loiseleux}}, \bibinfo {author} {\bibfnamefont {J.}~\bibnamefont {Chomaz}}, \
  and\ \bibinfo {author} {\bibfnamefont {P.}~\bibnamefont {Huerre}},\
  }\href@noop {} {\bibfield  {journal} {\bibinfo  {journal} {Physics of
  Fluids}\ }\textbf {\bibinfo {volume} {10}},\ \bibinfo {pages} {1120}
  (\bibinfo {year} {1998})}\BibitemShut {NoStop}%
\bibitem [{\citenamefont {Hoff}\ \emph {et~al.}(2016)\citenamefont {Hoff},
  \citenamefont {Harlander},\ and\ \citenamefont
  {Egbers}}]{hoff_harlander_egbers_2016}%
  \BibitemOpen
  \bibfield  {author} {\bibinfo {author} {\bibfnamefont {M.}~\bibnamefont
  {Hoff}}, \bibinfo {author} {\bibfnamefont {U.}~\bibnamefont {Harlander}}, \
  and\ \bibinfo {author} {\bibfnamefont {C.}~\bibnamefont {Egbers}},\ }\href
  {\doibase 10.1017/jfm.2015.743} {\bibfield  {journal} {\bibinfo  {journal}
  {Journal of Fluid Mechanics}\ }\textbf {\bibinfo {volume} {789}},\ \bibinfo
  {pages} {589–616} (\bibinfo {year} {2016})}\BibitemShut {NoStop}%
\bibitem [{\citenamefont {Evans}\ and\ \citenamefont {Morriss}(1984)}]{EVANS}%
  \BibitemOpen
  \bibfield  {author} {\bibinfo {author} {\bibfnamefont {D.~J.}\ \bibnamefont
  {Evans}}\ and\ \bibinfo {author} {\bibfnamefont {O.}~\bibnamefont
  {Morriss}},\ }\href@noop {} {\bibfield  {journal} {\bibinfo  {journal}
  {Computer Physics Reports}\ }\textbf {\bibinfo {volume} {1}},\ \bibinfo
  {pages} {297 } (\bibinfo {year} {1984})}\BibitemShut {NoStop}%
\bibitem [{\citenamefont {Salin}\ and\ \citenamefont {Caillol}(2002)}]{Salin}%
  \BibitemOpen
  \bibfield  {author} {\bibinfo {author} {\bibfnamefont {G.}~\bibnamefont
  {Salin}}\ and\ \bibinfo {author} {\bibfnamefont {J.-M.}\ \bibnamefont
  {Caillol}},\ }\href@noop {} {\bibfield  {journal} {\bibinfo  {journal} {Phys.
  Rev. Lett.}\ }\textbf {\bibinfo {volume} {88}},\ \bibinfo {pages} {065002}
  (\bibinfo {year} {2002})}\BibitemShut {NoStop}%
\bibitem [{\citenamefont {Khrapak}(2016)}]{khrapak2016relations}%
  \BibitemOpen
  \bibfield  {author} {\bibinfo {author} {\bibfnamefont {S.~A.}\ \bibnamefont
  {Khrapak}},\ }\href@noop {} {\bibfield  {journal} {\bibinfo  {journal}
  {Physics of Plasmas}\ }\textbf {\bibinfo {volume} {23}},\ \bibinfo {pages}
  {024504} (\bibinfo {year} {2016})}\BibitemShut {NoStop}%
\bibitem [{\citenamefont {Kundu}\ \emph {et~al.}(2015)\citenamefont {Kundu},
  \citenamefont {Cohen},\ and\ \citenamefont {Dowling}}]{Kundu:book}%
  \BibitemOpen
  \bibfield  {author} {\bibinfo {author} {\bibfnamefont {P.~K.}\ \bibnamefont
  {Kundu}}, \bibinfo {author} {\bibfnamefont {I.~M.}\ \bibnamefont {Cohen}}, \
  and\ \bibinfo {author} {\bibfnamefont {D.~R.}\ \bibnamefont {Dowling}},\
  }\href@noop {} {\emph {\bibinfo {title} {{Fluid Mechanics}}}},\ \bibinfo
  {edition} {6th}\ ed.\ (\bibinfo  {publisher} {Academic Press},\ \bibinfo
  {address} {San Diego},\ \bibinfo {year} {2015})\BibitemShut {NoStop}%
\end{thebibliography}
\providecommand{\noopsort}[1]{}\providecommand{\singleletter}[1]{#1}%
\end{document}